\documentclass[english,conference]{IEEEtran}
\usepackage[T1]{fontenc}
\usepackage[latin9]{inputenc}
\usepackage{amsmath}
\usepackage{graphicx}
\usepackage{amssymb}

\makeatletter
\usepackage{amsfonts}

\makeatother

\usepackage{babel}

\begin{document}

\title{PMI-based MIMO OFDM PHY Integrated Key Exchange (P-MOPI) Scheme}
\author{\IEEEauthorblockN{Pang-Chang Lan\dag, Chih-Yao Wu\dag, Chia-Han Lee\ddag, Ping-Cheng Yeh\dag, and Chen-Mou Cheng\dag\ddag}
\bigskip
\IEEEauthorblockA{\dag Department of Electrical Engineering and\\
Graduate Institute of Communication Engineering\\
National Taiwan University\\
Taipei, Taiwan}
\bigskip
\IEEEauthorblockA{\ddag Research Center for Information Technology Innovation\\
 Academia Sinica\\
Taipei, Taiwan}}
\date{\empty}
\maketitle
\begin{abstract}
In~\cite{Cheng2010}, we have proposed the MIMO-OFDM PHY integrated
(MOPI) scheme for achieving physical-layer security in practice without using any cryptographic ciphers. The MOPI scheme
uses channel sounding and physical-layer network coding (PNC) to prevent
eavesdroppers from learning the channel state information (CSI). Nevertheless,
due to the use of multiple antennas for PNC at transmitter and beamforming at
receiver, it is not possible to
have spatial multiplexing nor use space-time codes in our previous
MOPI scheme. In this paper, we propose a variant of the MOPI
scheme, called P-MOPI, that works with a cryptographic cipher and utilizes precoding matrix index (PMI) as an
efficient key-exchange mechanism. With channel sounding, the PMI is
only known between the transmitter and the legal receiver. The shared
key can then be used, e.g., as the seed to generate pseudo random bit sequences
for securing subsequent transmissions using a stream cipher. By applying the same
techniques at independent subcarriers of the OFDM system, the P-MOPI
scheme easily allows two communicating parties to exchange over 100 secret bits. As a result, not
only secure communication but also the MIMO gain can be guaranteed
by using the P-MOPI scheme.
\end{abstract}

\section{Introduction}
In his seminal work, Wyner showed that, by information-theoretic arguments,
it is possible to achieve communication confidentiality by exploiting
the spatial diversity of wireless channels \cite{Wyner1975}. Wyner's
theoretical investigation has aroused many different proposals, such
as \cite{Li2006a,Jorgensen2007,Goel2008,Kobayashi2008,Lakshmanan2008}.
While most of the existing physical-layer (PHY) security works focus on information
theoretical approaches that are difficult to implement in reality,
our recent work \cite{Cheng2010} is one of the pioneering works that
propose practical schemes for PHY security. In the
work, a MIMO-OFDM PHY integrated (MOPI) scheme has been proposed to
provide communication confidentiality without using any cryptographic ciphers in wireless networks. By the
use of channel sounding, MOPI prevents an eavesdropper from learning
the channel state information (CSI) of the channel between the eavesdropper
and the transmitting node from the preambles or pilot tones. A bit-interleaved
coded-modulation (BICM)-based physical layer network coding (PNC)
scheme has been proposed such that an eavesdropper, due to lack of
CSI, will suffer from very high bit error rate (BER) in decoding.
The MOPI scheme has been shown to provide excellent security, forcing
the eavesdropper to have large estimation error even if blind channel
estimation is used.  The computational complexity is also prohibitively expensive  if the eavesdropper resorts
to brute-force search to recover the CSI.

Although the MOPI scheme is promising in providing realistic PHY security, it has a serious drawback. Since the
multiple antennas at the transmitter are used for PNC and the multiple
antennas at the receiver are used for beamforming, it is impossible
for this MOPI scheme to have spatial multiplexing nor use space-time codes.
This limits the capability of the MIMO system. In this paper, 
a variant of the MOPI scheme---called P-MOPI---based on the precoding matrix
index (PMI) is proposed. The precoding matrix index, commonly applied
in MIMO systems nowadays, is used as the secret key. It is well known
that the MIMO system performance can be enhanced by precoding at the
transmitter, i.e., multiplying the signal vector by a matrix before
transmission. With the optimal precoding at the transmitter, the MIMO channel
can be transformed into several parallel subchannels, and the channel
capacity can be achieved. Typically there is a universal codebook
that consists of a finite number of precoding matrices. Due to different
channel realizations between the transmitter-legal receiver and the
transmitter-eavesdropper pairs, the precoding matrix is only known
between the transmitter and the legal receiver, so the precoding matrix
indices can therefore be used as keys. With the proposed efficient
key-exchange mechanism, the shared secret key, easily over 100 bits
long, can be used, e.g., as the seed to generate pseudo random bit sequences, and
secured MIMO communications can then be achieved, by using a stream cipher.
This is the fundamental difference between MOPI and P-MOPI despite
their otherwise striking similarity: while MOPI is designed to
\emph{replace} a cryptographic cipher and encrypts messages at
physical layer, P-MOPI is designed to \emph{work} with one---it merely
uses physics to establish the shared key between the communicating
parties.

The rest of this paper is organized as follows.
The relevant background information is reviewed in Section~\ref{sec:backgrounds}.
The scenario considered
is described in Section \ref{sec:System-Setup}. The PMI-based
key-exchange scheme is described in Section \ref{sec:PMI-based-Key-Distribution}.
The performance of the proposed P-MOPI scheme is evaluated using computer
simulation, with the results presented and discussed in Section \ref{sec:Performance-Evaluation}.
Conclusions are addressed in Section \ref{sec:Conclusions}.

\section{Backgrounds}
\label{sec:backgrounds}

\subsection{MIMO Systems with Precoding}

First let us review how the precoding matrix is used in the MIMO system.
Alice first sends out a reference signal for Bob to estimate the channel
matrix $\mathbf{H}^{AB}$ and decides the optimal precoding matrix.
Note that the channel here stands for the channel on a subcarrier
or on certain subcarriers of OFDM, and the index of subcarrier is
omitted for simplicity. In practical situations, in order to reduce
the complexity and the feedback overhead, a universal codebook $\mathcal{F}$ that
consists of a finite number of precoding matrices is used. Each precoding
matrix in the codebook has an index called precoding matrix index
(PMI).

Consider the following MIMO channel capacity formula \begin{align}
\mathcal{C}_{\mathbf{H},\mathbf{F}}=\log_{2}\det[\mathbf{I}_{n}+\frac{E_{s}}{n_{s}\sigma^{2}}\mathbf{F}^{\dagger}\mathbf{H}^{\dagger}\mathbf{H}\mathbf{F}],\label{equation 1}\end{align}
where $\mathbf{I}_{n}$ is the identity matrix with $n$ denoting
the minimum number of antennas at Alice and Bob, $E_{s}$ is the symbol
energy, $n_{s}$ is the noise power, $\sigma^{2}$ is the noise variance,
$\dagger$ means the Hermitian, $\mathbf{F}$ is the precoding matrix,
and $\mathbf{H}$ is the channel. Bob finds the precoding matrix and
its corresponding PMI from the codebook that maximizes the channel
capacity. Mathematically, \begin{align}
\hat{\mathbf{F}}=\underset{\mathbf{F} \in \mathcal{F}}{\operatorname{argmax}}\ \mathcal{C}_{\mathbf{H},\mathbf{F}},\label{equation 2}\end{align}
where $\hat{\mathbf{F}}$ is the best precoding matrix from the codebook
$\mathcal{F}$. We denote the PMI associated with  $\hat{\mathbf{F}}$  by $i_{\mathrm{PMI}}$.

\subsection{Stream Ciphers}

A stream cipher is a symmetric-key cipher based on the idea of Shannon's
one-time pads. Typically, a stream cipher generates a sequence of
a pseudorandom bit stream called the \emph{keystream}, which is XOR'ed
with the plaintext. It is this bit-by-bit way of encryption that gives
the name ``stream cipher.''

A perhaps simplest form of a stream cipher is a linear feedback shift
register (LFSR) with its output fed into a nonlinear filter function
to generate the keystream. Note that the nonlinear filter function
is of essential importance here; otherwise the structure of an $n$-bit
LFSR can easily be recovered using the Berlekamp-Massey algorithm
given 2$n$ keystream bits. Such a nonlinear filter, if well-designed,
can resist most attacks if the number of leaked keystream bits is
relatively small compared with the product of the LFSR width and the
algebraic degree of the nonlinear filter. The interested reader is
referred to~\cite{Canteaut2005} for further detail in filter design.
It is also possible to construct other kinds of LFSR-based stream
ciphers. One such example is Trivium, a combiner generator type of
stream cipher, which has a security level of $2^{80}$
and can be implemented with slightly more than two thousand NAND gate
equivalents~\cite{Mentens}.

It is also possible to construct a stream cipher using a standard
or light-weight block cipher. For example, KATAN and KTANTAN belong
to a family of small and efficient block ciphers. The family is based
on LFSR, and they share the same key size of 80 bits. KTANTAN is smaller
than KATAN, but the key is burnt into the device and hence can not
be changed. The smallest cipher in the family can be implemented in
less than 500 gates in 0.13 $\mu m$ CMOS technology,
while achieving encryption speed of 12.5 kbits/sec. More details about
KATAN and KTANTAN can be found in~\cite{Canniere2009}.

\begin{figure}[!t]
\begin{centering}
\includegraphics[width=2.5in]{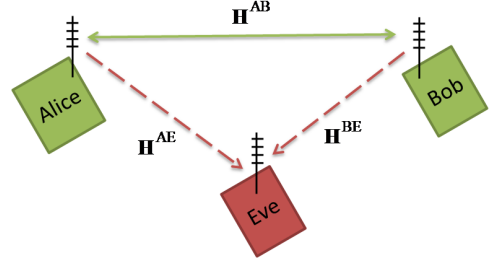} \caption{System Model.}
\label{Flo:setup}
\par\end{centering}
\end{figure}

\section{System Setup \label{sec:System-Setup}}

The system model, shown in Fig. \ref{Flo:setup}, consists of three
nodes, Alice, Bob, and Eve, and three wireless MIMO channels: $\mathbf{H}^{AB}$,
$\mathbf{H}^{AE}$, and $\mathbf{H}^{BE}$. The source node, Alice,
wants to transmit confidential messages to the destination node, Bob,
through $\mathbf{H}^{AB}$. Due to the broadcast nature of wireless
channels, these messages will be overheard by the eavesdropper, Eve,
through $\mathbf{H}^{AE}$. If Bob transmits some signals to Alice,
those signals will also be overheard by Eve through $\mathbf{H}^{BE}$.
It is assumed that the channel between Alice and Bob is symmetric,
i.e., $\mathbf{H}^{AB}=\mathbf{H}^{BA}$, and the channel realizations
of $\mathbf{H}^{AB}$, $\mathbf{H}^{AE}$, and $\mathbf{H}^{BE}$
are independent to each other. Alice is assumed to have four antennas
while Bob is assumed to have two antennas and both Alice and Bob use the OFDM technique
during transmission, whereas Eve can have an arbitrary number of antennas.
%A block-fading channel is assumed, so the channel realization is considered
%constant within one codeword.

As will be described in details later, a universal codebook containing
precoding matrices and PMI's for precoding is available to Alice,
Bob, and Eve, and the MIMO channel capacity function for PMI estimation
is known by Eve.

\section{The P-MOPI Schemes \label{sec:PMI-based-Key-Distribution}}

In the original MOPI scheme \cite{Cheng2010}, the multiple antennas
at the transmitter are used for PNC and the multiple antennas at the
receiver are used for beamforming. This provides highly secure wireless 
communications. Yet, the  scheme sacrifices the MIMO capability of using spatial
multiplexing or space-time codes. In order to use antennas in a more
efficient manner, we propose the P-MOPI scheme in this paper, which allows
Alice and Bob to use PMI to exchange keys for subsequent use in cryptographic ciphers to secure their communications.

\subsection{P-MOPI}

In a typical MIMO system with precoding, Alice acquires the PMI
via the feedback from Bob. Eve can easily detect the PMI through eavesdropping.
But what if the PMI is not fed back to Alice, and instead, Bob sends
the same reference signal to Alice? Under the assumption of channel
reciprocity ($\mathbf{H}^{AB}=\mathbf{H}^{BA}$), Alice is able to
compute the PMI that is the same as Bob's. At Eve's side, without
the feedback, she is unable to figure out the PMI since the channel
$\mathbf{H}^{AE}$ is different from $\mathbf{H}^{AB}$ (and $\mathbf{H}^{BA}$).
Now the PMI, only shared between Alice and Eve, can be used as a secret
key.

The typical size of PMI is $2$ to $6$ bits long depending on the
number of MIMO antennas. Let us assume it is $4$ bits. Through our scheme,
Alice and Bob can share $4$ secret bits over \emph{one} subcarrier.
In OFDM systems, the independent fading realizations between subcarriers
or subbands lead to abundant generation of independent PMI's.
%Fig. \ref{fig:freqcorrelation} shows the correlation between fading
%realizations of different subcarriers.
%\begin{figure}
%  \begin{center}
%    \includegraphics[width=.45\textwidth]{figure/freqcorrelation}
%    \caption{Correlation coefficients between fading realizations of
%      different subcarriers}
% \label{fig:freqcorrelation}
%  \end{center}
%  \end{figure}
%In general, two subcarriers that are more than ~KHz apart are
%almost uncorrelated, where their correlated coefficient drops below 0.2.
As long as the whole channel can be divided into more than $25$ subbands with nearly uncorrelated
fading which are easy to acquire in general, our scheme enables Alice and Bob to share over $100$ secret bits. The
secret key can then be used to generate a random sequence for data
to be transmitted securely using a stream cipher. With a key of size over $100$ secret bits, a security level way above
$\mathcal{O}\left(2^{100}\right)$ (meaning Eve should try at least
about $\mathcal{O}\left(2^{100}\right)$ times to approach her best
performance) can be achieved, which is the usual strength requirement
for a cryptographic cipher.

The steps of the proposed scheme are summarized below:
\begin{enumerate}
\item Alice transmits a reference signal to Bob to let Bob make channel estimation.
\item Bob estimates the channel on a single subcarrier or a subband (which consists
of several subcarriers depending on the channel coherence bandwidth).
Channel realization $\mathbf{H}^{AB}$ is acquired at Bob's side.
\item Bob conducts the corresponding precoding matrix $\hat{\mathbf{F}}_{\mathrm{Bob}}$
for $\mathbf{H}^{AB}$ by finding $\underset{\mathbf{F}}{\operatorname{argmax}}\ \mathcal{C}_{\mathbf{H},\mathbf{F}}$.
He regards the PMI $i_{\mathrm{PMI,Bob}}$ of the precoding matrix
$\hat{\mathbf{F}}_{\mathrm{Bob}}$ as a key and put it into his key
set $\mathcal{K}_{\mathrm{Bob}}$.
\item During the next time slot, Bob sends a sounding signal to Alice. Alice
acquires the corresponding precoding matrix $\hat{\mathbf{F}}_{\mathrm{Alice}}$
for $\mathbf{H}^{BA}$ for every subcarrier. Alice then puts $i_{\mathrm{PMI,Alice}}$
into its key set $\mathcal{K}_{\mathrm{Alice}}$. Since the channel
reciprocity holds, $\hat{\mathbf{F}}_{\mathrm{Bob}}$ and $\hat{\mathbf{F}}_{\mathrm{Alice}}$
are the same, and so are $\mathcal{K}_{\mathrm{Bob}}$ and $\mathcal{K}_{\mathrm{Alice}}$.
\item Alice and Bob may drop out-of-date keys to make sure $\mathcal{K}_{\mathrm{Bob}}=\mathcal{K}_{\mathrm{Alice}}$
at any time.
\item Alice uses a stream cipher to encrypt data with the key set $\mathcal{K}_{\mathrm{Alice}}$.
Afterwards, Alice transmits the encrypted data to Bob and Bob decrypts
the data using its own key set $\mathcal{K}_{\mathrm{Bob}}$. During
the transmission, precoding is applied in order to achieve better
MIMO performance.
\item (Optional rekeying) In rare situations where
  $\mathcal{K}_{\mathrm{Bob}}\neq\mathcal{K}_{\mathrm{Alice}}$, Alice
  and Bob need to rekey by going to Step 1.  Such a mismatched key can
  be detected, e.g., as follows.  Alice first picks a random number
  $X$ and transmits the encryption of $X$ under
  $\mathcal{K}_{\mathrm{Alice}}$, along with the SHA-256 digest of $X$
  in \emph{plaintext}.  At the receiving end, Bob decrypts using
  $\mathcal{K}_{\mathrm{Bob}}$, calculates its SHA-256 digest, checks
  to see if it matches the received digest, and declares a rekeying if
  there is a mismatch.
\end{enumerate}

\subsection{P-MOPI for slow-varying channel}

In the P-MOPI scheme, PMI is used as the key. In the slow-varying
channel, the optimal precoding matrix will stay similar, so the PMI
(and the key) will be the same for a long period of time. From the security
point of view, keys should be changed frequently, so that means the
basic P-MOPI scheme proposed earlier only works well under the fast-varying
channel. For the slow-varying channel, we need to make a revision,
as proposed below.
\begin{enumerate}
\item Alice transmits the reference signal $\mathbf{s}$ regularly in order
to update her precoding matrix to match the latest channel condition.
\item If the new PMI is different from the previous PMI, Bob replies with
the reference signal $\mathbf{r}$; if the new PMI is the same as
the previous PMI, Bob first replies a flag bit to inform Alice the channel is static,
and then sends a rotated reference signal $\mathbf{Ur}$ immediately,
where $\mathbf{U}$ is a randomly generated unitary matrix.
\item The flag bit is for Alice to remain the same \emph{precoding} for the static channel
, and the rotated reference signal $\mathbf{Ur}$ is for her to obtain
the \emph{key} through PMI estimation. Since
Bob knows $\mathbf{U}$, he obtains the PMI as well.
\item Alice then uses the key for stream cipher as before.
\item Since Alice still transmits the normal reference signal $\mathbf{s}$ regularly,
Bob can estimate $\mathbf{H}^{AB}$ all the time. If the PMI for \emph{precoding} is not the same as the previous PMI,
it means the channel becomes dynamic. Then Bob transmits another flag bit to inform Alice to change back to normal P-MOPI scheme without $\mathbf{U}$.
\end{enumerate}
Notice that in order to estimate the PMI for the reference signal
$\mathbf{Ur}$, Alice and Bob need to perform the following operation
(instead of Eqn. (\ref{equation 2})): \begin{align}
\hat{\mathbf{F}}=\underset{\mathbf{F}\in\mathcal{F}}{\operatorname{argmax}}\ \bar{\mathcal{C}}_{\mathbf{H},\mathbf{F}},\end{align}
where

\begin{eqnarray}
\bar{\mathcal{C}}_{\mathbf{H},\mathbf{F}} & = & \log_{2}\det[\mathbf{I}_{n}+\frac{E_{s}}{n_{s}\sigma^{2}}\mathbf{F}^{\dagger}\left(\mathbf{H\mathbf{U}}\right)^{\dagger}\left(\mathbf{H}\mathbf{U}\right)\mathbf{F}],\\
 & = & \log\det[\mathbf{I}_{n}+\frac{E_{s}}{n_{s}\sigma^{2}}\mathbf{\bar{F}}^{\dagger}\mathbf{H}^{\dagger}\mathbf{H}\mathbf{\mathbf{\bar{F}}}],\label{eq:cu}\end{eqnarray}
where $\mathbf{\bar{F}}=\mathbf{U}\mathbf{F}$. This means that Alice
tries to find the best precoding matrix based on a modified channel
capacity function.

The purpose of introducing $\mathbf{U}$ is to generate a new key
when the channel does not change much. Apparently Eve cannot acquire any information about $\mathbf{U}$.
Moreover, the PMI is obtained through an optimization process, which is an nonlinear
process. Without knowing the channel $\mathbf{H}$,
 it is hopeless for Eve to obtain the correct PMI even when the codebook and the MIMO
channel capacity function is known publicly. However, we should consider the influence of $\mathbf{U}$.
For Alice, the estimation effective channel is $(\mathbf{H_{AB}}+\mathbf{n})\mathbf{U} = \mathbf{H_{AB}}\mathbf{U} +\mathbf{n}\mathbf{U}$, where $\mathbf{n}$ is denoted as white gaussian noise. For Bob, the estimation effective channel is $\mathbf{H_{AB}}\mathbf{U}+\mathbf{n}$. If $\mathbf{U}$ is chosen as unitary matrix, the estimation error of Alice has the
same statistics compared to Bob's estimation error. It is obvious that there is no noise enhancement with the multiplication of $\mathbf{U}$.

Compared to the basic P-MOPI scheme, the P-MOPI scheme for the slow-varying
channel needs to transmit one extra flag bit and requires
Alice and Bob to do extra computation for finding PMI. Nevertheless,
the security is highly improved.

\section{Scheme Evaluation \label{sec:Performance-Evaluation}}

The success of the proposed P-MOPI relies on the following factors:
channel coherence time and channel coherence bandwidth. Channel coherence
time determines whether Alice and Bob can obtain the same PMI and
channel coherence bandwidth decides how many independent channels
can be obtained. The more available independent channels, the more
secret bits can be generated. In this section, we will evaluate the
feasibility of the proposed P-MOPI scheme.

We take the simulation set up generally used in popular 4G standard - Long Term Evolution (LTE) \cite{TR25996,TR36814} to evaluate P-MOPI scheme. The detailed simulation parameters are provided in the following tables.
\begin{tabular}[t]{ll}
\hline
Simulation Setup \\
\hline
Channel model & SCME channel model \\
MIMO system & $4\times 2$ single user MIMO \\
Subcarrier bandwidth & $15$ kHz\\
Total bandwidth & $20$ MHz\\
Center frequency & $2$ GHz\\
SCME scenario & Urban macro\\
Bob's velocity & $0$, $3$, $10$ (km/hr)\\
Codebook & 4-bit Householder codebook \\
\hline
\end{tabular}\\\\

\subsection{Channel coherence bandwidth}
\begin{figure}[!t]
\begin{centering}
\includegraphics[width=3.5in]{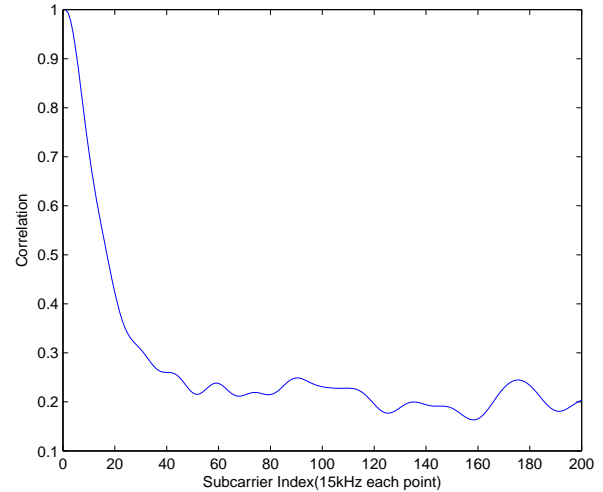} \caption{Channel correlation among the subcarriers under noiseless Spatial Channel Modeling Extended (SCME) urban-macro channel model.}
\label{Flo:freqcorr}
\par\end{centering}
\end{figure}

The security of the communications mainly depends on the size of the key space, i.e. the number of available keys. Hence, it is desirable to know how many available keys can be generated by our P-MOPI scheme. Fig. \ref{Flo:freqcorr} shows the simulation result of the channel correlation among the subcarriers in the SCME urban-macro channel model. In general, a correlation below $0.5$ can be regarded as nearly uncorrelated. We can see that the correlation decreases to $0.5$ at about a $20$-subcarrier separation, which shows that the coherence bandwidth of the channel is about $20\times 15\text{k} = 300$~kHz. With total system bandwidth $20$~MHz, the number of independent PMI's can be acquired is $20\text{M}/ 300\text{k} \cong 66$ in one time slot. With the $4$-bit Householder codebook, $4\times66=264$ bits are generated as the cryptography key. It outperforms the conventional required security level of $80$-bits key block cipher.

\subsection{Probability of Alice and Bob obtaining the same key}
\begin{figure}[!t]
\begin{centering}
\includegraphics[width=3.5in]{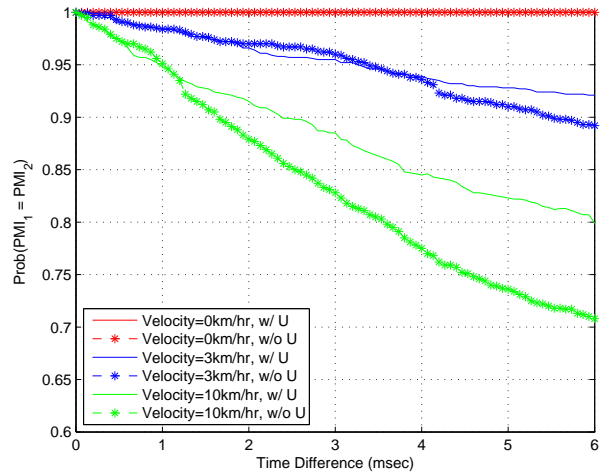} \caption{Probability of Alice and Bob obtaining the same PMI versus the time difference of channel esitmations under noiseless SCME urban-macro channel model.}
\label{Flo:timePMI}
\par\end{centering}
\end{figure}

In the wireless communications, the velocity of the mobile device significantly affects the channel coherence time, which determines the downlink/uplink switching duration in our P-MOPI scheme. In Fig. \ref{Flo:timePMI}, it depicts the probability of Alice and Bob obtaining the same PMI versus the time difference of their channel estimations. Note that the time difference is resulted from the delay of Bob sending the reference signal back to Alice. When Bob is not moving, the channel is static and the PMIs are always the same. With Bob moving at a speed of $3$~km/hr, the probability can be held above $0.98$ for $1$~ms. With Bob moving at $10$km/hr, the probability remains above $0.95$ for $1$~ms. Remember that the random rotation unitary matrix $\mathbf{U}$ is used for the slow-varying channel. From the figure, the existence of $\mathbf{U}$ does not decrease the performance. It even increases the probability when Alice's and Bob's channels are not closely correlated. The concept is that if two channel realizations are separated from each other, the multiplication of $\mathbf{U}$ provides them the similar basis. So the two channel realizations will become closer after the rotation by $\mathbf{U}$.

\subsection{Influence of channel estimation error}
\begin{figure}
  \begin{center}
    \includegraphics[width=.45\textwidth]{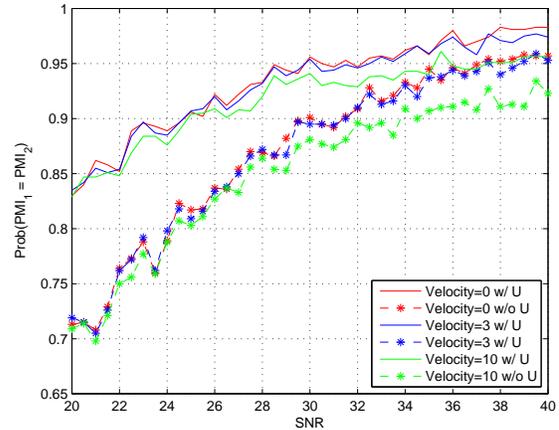}
    \caption{PMI reciprocity probability versus SNR under SCME urban-macro channel model}
  \label{fig:snr}
   \end{center}
\end{figure}

In this subsection, we illustrate the influences of the channel estimation error in Fig. \ref{fig:snr}. The estimation error is modeled as a Gaussian noise. The SNR is defined as the ratio of reference signal power to the noise variance. From the figure, at a moderate SNR like $40$dB, the influence of estimation error can be neglected. Especially, the reference signal power is usually set large in order to ensure the correctness of channel estimation. And we can see that with $\mathbf{U}$, the probability of both PMI's being the same is significantly increased. 

\subsection{Rekeying}

The P-MOPI scheme requires channel reciprocity in order for achieving
key exchange between Alice and Bob. Even though the
channel is reciprocal, sometimes the noise or interference in the
channel can be so high, and the received reference signals at Alice and
Bob are severely distorted such that $i_{\mathrm{PMI,Alice}}\neq
i_{\mathrm{PMI,Bob}}$.  In these cases, Alice and Bob can not
establish a shared secret key and will need to rekey by
restarting the whole procedure.

Fig. \ref{Flo:timePMI} shows the probability that the channel
between Alice and Bob is indeed reciprocal in high-SNR or nearly
noiseless situations. We can see that as long as the time difference between Alice's and
Bob's channel soundings is within 1~ms, the probability that
$i_{\mathrm{PMI,Alice}}=i_{\mathrm{PMI,Bob}}$ is very close to $1$, resulting in low chance of rekeying. However, for channels
with higher mobility, the probability drops and the chance of rekeying gets higher. 

Although rekeying is not a fatal problem as Alice and Bob can restart
the keying process and will succeed in a small number of rounds with
high probability, frequent rekeying will result in performance
degradation, which happens when the channel varies too quickly or when
the mobility is too high.
Therefore, it is still desirable to incorporate channel coding with our P-MOPI in our future work to reduce the
chance of rekeying. This can be done by transmitting reference signals over contiguous subcarriers
and use the similarity of the channels experienced by these subcarriers as source of redundancy. Another
approach would be to modify the P-MOPI design so that Bob can have some control on the PMI experienced by Alice instead 
of solely determined by the channel. If this can be done, channel coding can be applied. Such design is currently in progress.

%Fig. \ref{fig:snr} shows the same probability at a fixed time
%difference (1.33~ms) between Alice's and Bob's pilot signals.
%We can see that the rekeying probability is reasonably low, e.g., no
%more than 0.2 when the SNR is at least 15~dB.

\section{Concluding Remarks \label{sec:Conclusions}}

In this paper, we have proposed an efficient secret key exchange mechanism
for MIMO-OFDM systems. The proposed P-MOPI scheme utilizes the precoding
matrix indices as secret keys. The PMI is obtained by finding the
precoding matrix that maximizes the MIMO channel capacity function.
Due to independent channel realizations, the eavesdropper is unable
to learn the channel state information between the transmitter and
the legal receiver, resulting in the secure communication.

Two P-MOPI schemes were proposed for fast-varying and slow-varying
channels respectively. A random matrix is introduced in the P-MOPI
scheme for the slow-varying channel condition to update keys frequently.
The P-MOPI scheme, unlike the previous version of MOPI scheme, can
take full advantage of the multiple antennas through precoding. The
feasibility of the scheme has been evaluated through computer simulations.

Finally, we note that it is also possible to use a randomness
extractor such as a cryptographic hash function under appropriate
circumstances~\cite{fouque2008hmac} to extract shared secrets from the
PMIs of \emph{all} subcarriers instead of uncorrelated subcarriers, as
we have proposed and examined in this paper.  It warrants further
investigation as whether this will improve or worsen the rekeying
probability, an important trade-off between security and usability.

\section*{Acknowledgments}

\bibliographystyle{IEEEtran} \bibliographystyle{IEEEtran}
\bibliography{bib/list_P-MOPI}

\end{document}